\documentclass[reprint,
superscriptaddress,
 amsmath,amssymb,
floatfix,
]{revtex4-2}

\usepackage{graphicx}
\usepackage{dcolumn}
\usepackage{bm}
\usepackage[breaklinks=true,colorlinks=true,linkcolor=blue,urlcolor=blue,citecolor=blue]{hyperref}
\usepackage{epstopdf}
\usepackage{siunitx}
\DeclareSIUnit{\dBm}{dBm}
\usepackage{enumitem}
\usepackage{color}
\begin{document}


\title{Modelling spin-wave interference with electromagnetic leakage in micron-scaled spin-wave transducers}
\author{Felix Kohl}
\email{fkohl@rptu.de}
\affiliation{Fachbereich Physik and Landesforschungszentrum OPTIMAS,
Rheinland-Pfälzische Technische Universität Kaiserslautern-Landau,
D-67663 Kaiserslautern, Germany}

\author{Björn Heinz}
\affiliation{Fachbereich Physik and Landesforschungszentrum OPTIMAS,
Rheinland-Pfälzische Technische Universität Kaiserslautern-Landau,
D-67663 Kaiserslautern, Germany}

\author{Matthias Wagner}
\affiliation{Fachbereich Physik and Landesforschungszentrum OPTIMAS,
Rheinland-Pfälzische Technische Universität Kaiserslautern-Landau,
D-67663 Kaiserslautern, Germany}

\author{Christoph Adelmann}
\affiliation{imec, Kapeldreef 75, 3001 Heverlee, Belgium}

\author{Florin Ciubotaru}
\affiliation{imec, Kapeldreef 75, 3001 Heverlee, Belgium}

\author{Philipp Pirro}%
\affiliation{Fachbereich Physik and Landesforschungszentrum OPTIMAS,
Rheinland-Pfälzische Technische Universität Kaiserslautern-Landau,
D-67663 Kaiserslautern, Germany}

\date{\today}

\begin{abstract}
Utilization of spin-wave transducers for radio-frequency signal processing provides significant potential due to intrinsic tunability, scalability and nonlinearity. However, signal interference with electromagnetic leakage can diminish their operation and functionality. Here, we experimentally identify the electromagnetic crosstalk (EM) as a major source of interference induced distortion and provide a simple analytic model to predict the impact on device operation. The results are in good agreement with the experimental observation. In addition, we test multiple transducer geometries to identify operational regimes and minimize the EM impact. Finally, the effect of nonlinear device operation on the interference is addressed, which is of relevance for the exploitation of the spin-waves intrinsic nonlinear traits.
\end{abstract}

\maketitle

\section{Introduction}
In recent years, magnonics has been extended from fundamental research  \cite{Schweizer,Roadmap2024,Pirro2021} to an increased focus on practical technological application  of spin waves \cite{Roadmap2024,Cocconcelli,Pirro2021,QiRepeater,QiCoupler,Levchenko}. In particular, due to their tunability, scalability and accessible frequency range \cite{Levchenko}, spin waves offer many promising traits for next-generation signal processing and logic systems \cite{Roadmap2024, LiYi, Davidkova, DevolderUni,QiCoupler}. Accordingly, several different approaches and prototypes for various use cases have been demonstrated in the literature in the recent past \cite{Breitbach2024, Qi2024,Papp2021,DuFilter,HanMagnonicDevices}. Many of these approaches are based on the inductive excitation and detection of spin waves using microstructured antennas on magnetic materials \cite{Vlaminck2010, Levchenko,DaiOctave,DuFilter}, especially YIG with its very low spin-wave attenuation \cite{Dubs2017, Qin}. 
Although significant progress has been made in improving the efficiency of spin-wave transducers \cite{Ich}, it should be noted that the spin-wave antennas also couple inductively via the RF electric fields, giving rise to electromagnetic crosstalk (EM) \cite{GreilSecondary,Bunea}. The resulting signal modulation, often referred to as passband ripples, caused by this additional coupling have been observed in many studies, but have often been neglected for not being part of the research question \cite{Davidkova,SongSingle,Schmoll}. However, such distortions may impair device operation and signal integrity, emphasizing a need to investigate their cause and effects.
In this paper, we present a simple modeling approach to describe and predict the influence and effects of EM on signal transmission in micron-scaled spin-wave transducers. It is demonstrated that the interference of the spin-wave signal and EM leads to passband ripples and modulations in group delay of the RF output.  Relevant parameters influencing the modulation frequency, i.e. spin-wave dispersion and antenna spacing, are identified. Moreover, it is shown that the magnitude of these modulations depends significantly on the relative strength of the spin-wave signal and the EM. In the experiments, it is shown that the amplitude ratio and consequently the occurrence of passband ripples depends on the antenna layout and operation power level and that their influence can be minimized by careful design of the transducers depending on their respective use case.

\section{Theoretical background}

In a conventional spin-wave transducer, the dynamic Oersted fields of the driving antenna excite spin waves in an adjacent magnetic material. These spin waves then propagate through the magnet and eventually reach the detection antenna. There, the dynamic magnetic fields of the spin waves induce an oscillating voltage in the antenna \cite{Ich,Levchenko,Bailleul2003, Vlaminck2010}. 
A simple analytical model for calculating this coupling is given in \cite{Vanderveken}. According to this model, the antenna's self-inductance $L_\text{m}$ induced by the spin waves driven underneath and the mutual inductance $\mathcal{M}$ mediated by the propagating spin waves can be expressed as

\begin{equation}\label{equ:SelfInductance}
    L_\text{m} =\mu_0 tl\int\textbf{h}_a^T(k)\chi \big(\omega(k)\big) \textbf{h}_a^*(k)\frac{dk}{2\pi},
\end{equation}

\begin{equation}\label{equ:MutualInductance}
    \mathcal{M} = \mu_0 tl\int\text{e}^{ikD-\frac{2\pi D}{\delta(k)}}\textbf{h}_a^T(k)\chi \big(\omega(k)\big)\textbf{h}_a^*(k)\frac{dk}{2\pi},
\end{equation}

\begin{figure}[htb]
\includegraphics{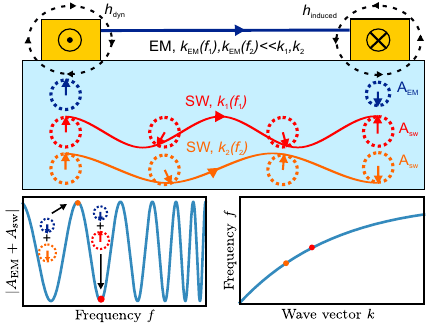}
\caption{ Schematic illustration of signal transmission by EM and spin-waves. The phase accumulation by the spin-wave is strongly dispersive. The resulting interference of both complex signal amplitudes is depicted in the left sub-panel. The right sub-panel shows the corresponding spin-wave dispersion relation.
\label{fig:SchematicInt}
}
\end{figure}

for an antenna of length $l$ on a magnet of thickness $t$ and an antenna pair of distance $D$ and the spin-wave decay length $\delta(k)$. In the expressions, $\chi \big(\omega(k)\big)$ represents the frequency-dependent dynamic susceptibility of the magnet and $\textbf{h}_a^*(k)$ the wave-vector-dependent Fourier amplitude of the dynamic antenna field. The resulting spectrum of frequency-dependent spin-wave transmission is described, e.g. in \cite{Ich}. Comparison of the two expressions directly shows that they differ only by a complex phase term in the integral, which describes the phase accumulation and attenuation of the spin waves as they propagate to the detection antenna.\\
In the experiment, usual spin-wave wavelengths are on the order of a few micrometers, well below the wavelength of microwave radiation at the same frequency which lies in the GHz range. With an antenna distance of $D=\SI{50}{\micro\metre}$, a considerable spin-wave phase accumulation is therefore expected.
In addition, the driving antenna also emits an electromagnetic near-field towards the detection antenna, which, on the contrary, has a negligible phase accumulation for such small distances. 
According to the induction law, this electromagnetic crosstalk (EM) also induces a voltage in the detection antenna, thus a direct superposition with the spin-wave contribution is measured. Since the accumulated spin wave phase depends on the wave vector $k(f)$, the relative phase between the two voltages varies with frequency. Accordingly, as the frequency is swept, points of constructive (orange in Fig. \ref{fig:SchematicInt}) and destructive interference (red in Fig. \ref{fig:SchematicInt}) occur, resulting in a modulation of the measured signal magnitude.
The lower right inset of Fig. \ref{fig:SchematicInt} shows a typical dispersion relation for YIG films with thickness in the micrometer range which is dominated by magnetostatic contributions (here: Damon-Eshbach configuration). Due to the non-linear spin-wave dispersion relation, the periodicity of the resulting oscillations of the signal magnitude is not constant, but itself frequency-dependent. For the steep dispersion at low wave vectors, the frequency steps with a relative phase shift of $\Delta \phi = \Delta k(f_1,f_2)D = 2\pi$, are large. While for larger wave vectors, the dispersion flattens noticeably, leading to a shorter oscillation period. Simultaneously, increasing propagation distance $D$ results in larger phase accumulation and shorter oscillation periods.

\begin{figure*}[ht]
\includegraphics{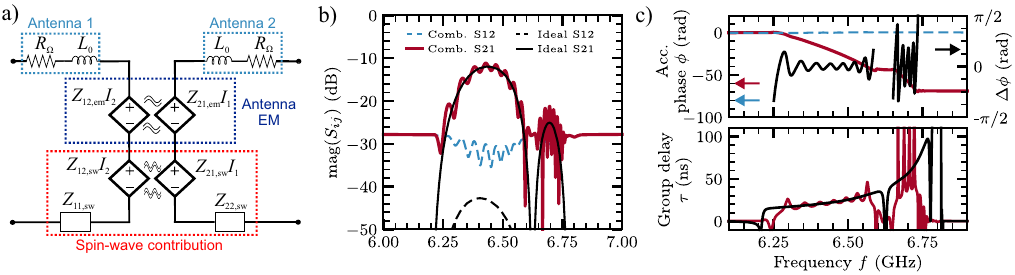}
\caption{ a)  Illustration of the combined network including the EM, adapted from \cite{RobertDesign}.
 b) Calculated transmission spectra for GS antenna on  $\SI{500}{nm}$ thick YIG film with $M_\text{s}=\SI{158.3}{\kilo\ampere\per\metre}$ and $A_\text{ex}=\SI{3.5}{\pico\joule\per\metre\cubed}$ in an external field of $\SI{143}{\milli\tesla}$ for forward  ($S_{21}$, solid lines)  and backward transmission  ($S_{12}$, dashed lines). Shown in black is the ideal case of zero crosstalk, colored lines refer to an EM of $\SI{2}{\ohm}$. c) Top panel: Unwrapped phase accumulation for $S_{21}$ (red). Shown in black is the difference to the pure spin-wave phase accumulation of $S_{21}$. Bottom panel: Group delay for $S_{21}$ with (red) and without (black) EM. \label{fig:CalcRipples} }
\end{figure*}

To model the superposition, we adapt the equivalent circuit for the two-port network from \cite{RobertDesign, Ich},  as shown in Fig. \ref{fig:CalcRipples} a). It describes the two-port network as a series connection of the electrical properties of the antennas including the inductive coupling via the EM and the contributions due to the spin-wave excitation (see Eqs. \ref{equ:SelfInductance}, \ref{equ:MutualInductance}). Using this adapted circuit diagram, the scattering matrix can be calculated, as shown in the Supplemental material. Exemplary, a pair of GS antennas with a spacing of $\SI{50}{\micro\metre}$ is considered, as it will be compared to the experiment in  section \ref{sec:Gating}. The antenna dimensions and their respective excitation efficiencies are presented in the supplementary material. An external field of $\SI{143}{\milli\tesla}$ is assumed and the impedance of the EM is set to be frequency-independent $\SI{2}{\ohm}$. This value is close to the mutual impedances measured for the off-resonant background for the same antenna in the experiment. The calculated transmission spectra in forward and backward direction are shown in Fig. \ref{fig:CalcRipples} b). The black curves show the spectrum without crosstalk, while the colored curves include the EM. The pronounced difference between the forward direction $S_{21}$ (solid) and the backward direction $S_{12}$ (dashed) is solely due to the ellipticity of the spin waves and the antenna fields and not caused by the crosstalk effect \cite{Ich}. For both directions, inclusion of the EM induces a periodic modulation of the signal magnitude, referred to as passband ripple. Due to flattening of the dispersion relation (see Fig. \ref{fig:SchematicInt}), the modulation period decreases towards higher frequencies. The two transmission directions allow a discrimination between two modulation regimes: In the case of strong spin-wave transmission ($S_{21}$, red), the spin-wave signal is significantly larger than the crosstalk,  leading to a modulation of the  spin-wave envelope. In contrast, for weak spin-wave transmission ($S_{12}$, blue), EM dominates and is modulated by the spin waves.\\
The interference of both contributions not only modulates the magnitude of the signal but also affects  phase and  group delay of the signal. The latter is defined as the frequency derivative of the phase, providing a measure of the propagation time and phase coherence when passing the transducer:
\begin{equation}\label{equ:Groupdelay}
\tau = -\frac{\text{d}\phi(\omega)}{\text{d}\omega}
\end{equation}
Please note that due to the small transducer spacing compared to the EM wavelength, a significant phase is only accumulated when spin waves can be excited in the spin-wave pass bands. To illustrate the effects of the interference of EM and spin wave signal, Fig. \ref{fig:CalcRipples} c) displays the phase difference between the undisturbed and the modulated case for a strong spin-wave transmission within the spin-wave band. As in the magnitude, the phase also shows a periodic modulation that becomes stronger towards the edges of the spin-wave band (when the spin-wave transmission becomes weaker). The impact on the group delay for this case, due to its definition as the derivative of the phase to the frequency, is even more pronounced as highlighted in the bottom panel. The impact on variations of the EM level, antenna spacing and external bias field are discussed in the supplementary material. The presented model illustrates that the influence of electromagnetic crosstalk on the transmitted phase poses challenges for phase-sensitive applications. Following the description of the origins and implications of the superposition of both contributions, the anticipated effects are examined experimentally below.\\


\section{Experimental setup}
Measurements are performed using propagating spin-wave spectroscopy (PSWS) \cite{Bailleul2003, Vlaminck2010, Qin,Vanataka} using an electromagnet and a two-port VNA. The sample is contacted using standard $\SI{150}{\micro\metre}$-pitch ground-signal-ground (GSG) RF probes. Before the measurements, a short-open-load-through calibration is performed using an on-chip calibration substrate. 
An input RF power of $\SI{-25}{\dBm}$  was selected and tested in advance to ensure operation of the transducers in the regime of linear spin-wave excitation.\\
The structures under test are fabricated on $\SI{500}{\nano\metre}$ thick YIG films on a $\SI{500}{\micro\metre}$ thick GGG substrate.
Using UV lithography, the spin-wave antennas made from  $10/\SI{250}{\nano\metre}$ titanium/gold are fabricated onto a $\SI{50}{\nano\metre}$ $\text{SiO}_\text{2}$ film deposited onto the YIG film using chemical vapor deposition at $\SI{250}{\degreeCelsius}$. The spacer film is inserted to avoid direct contact of the metal antenna with the YIG.
Overall, three types of antennas are produced to investigate the influence of the antenna layout and feed lines on the occurrence of crosstalk (see Fig. \ref{fig:AntennaDependence} a)) : a ground-signal (GS) antenna, 
a ground-signal-ground (GSG) antenna, 
and a GSGS antenna (two periods of the GS design). For all antennas, the dimensions, around $\SI{2}{\micro\metre}$ for all lines, and excitation efficiencies are presented in the supplementary material.

\section{Results}
To disentangle the respective contributions of spin waves and EM, a method to filter the experimental data is required, which is  presented in the following. The impact of antenna shape and, additionally, the power dependency are considered thereafter.

\subsection{Timegated transmission spectra}\label{sec:Gating}

\begin{figure*}[htb]
\includegraphics{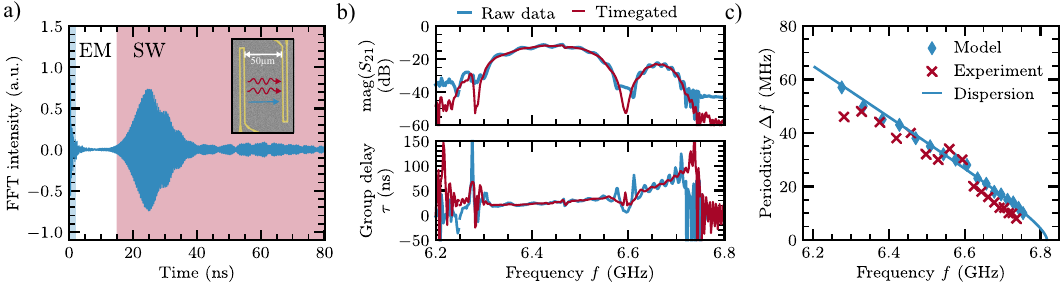}
\caption{a) Timetrace calculated from the frequency spectrum measured in b) for the GS antenna at an external field of $\SI{143}{\milli\tesla}$. Times corresponding to the EM and the spin-waves are shaded in color. The inset shows an scanning electron microscope image of the GS antenna pair including a schematic illustration of spin waves and EM. b) Measured transmission spectrum (top) and group delay (bottom) along with the spectrum obtained by setting the electric crosstalk to zero in a) and transformation back to the frequency domain.
c) Extracted periodicity of the  observed ripples in b) along with the periodicity calculated from the dispersion relation. Red crosses mark the periodicity extracted from the experiment on the GS antenna.\label{fig:Timegating}
}
\end{figure*}

In the experiment, the frequency-dependent transmission is  measured in the range from $\SI{2}{\giga\hertz}$ to $\SI{14}{\giga\hertz}$ in $\SI{2}{\mega\hertz}$ steps. The harmonic frequency spacing allows to convert the frequency spectrum into an artificial time-domain response following the method described in \cite{DevolderTimeGate}. In the following, a pair of GS antennas with $D=\SI{50}{\micro\metre}$ separation (see inset Fig. \ref{fig:Timegating} a)) are considered exemplary. Further details on the antenna are outlined in the supplementary material. The calculated time trace in Fig. \ref{fig:Timegating} a) corresponds to the output voltage in response to a Dirac impulse excitation at $t=0$ \cite{DevolderTimeGate}. This representation differs from the directly experimentally measured $S_{21}(f)$ continuous-wave (CW) excitation and offers valuable insight into the temporal evolution of the signal transmission. Owing to the method of Ref. \cite{DevolderTimeGate}, the resulting time trace is real-valued.\\
As can be seen in Fig. \ref{fig:Timegating} a), the largely different propagation velocities yield temporal separation between the signal contributions: Electrical crosstalk, propagating with the speed of light, arrives almost immediately at the output antenna, while spin waves - with group velocities of few kilometers per second - arrive only after about $\SI{20}{\nano\second}$. This separation allows one to filter out the influence the EM in post-processing, by excluding the first $\SI{2.5}{\nano\second}$ of the time trace (blue region in Fig. \ref{fig:Timegating} a)). The modified time trace is then transformed back into the frequency domain, resulting in a transmission spectrum without EM contribution. The same procedure, referred to as timegating in the following, applied to the $S_{12}$ trace is shown in the supplementary material.\\
Figure \ref{fig:Timegating} b) compares the raw and timegated spectra. The unfiltered signal shows strong passband ripples in the envelope of the spin waves. After time gating, these ripples are no longer present, confirming their origin in the interference of EM and spin-wave contributions. Accordingly, filtering the EM contribution also removes the oscillations in the group delay, as shown in the bottom panel. For a quantitative comparison, the oscillation periodicities extracted from the experiment and a calculation using the measured off-resonant EM are plotted in Fig. \ref{fig:Timegating} c) next to the expected values obtained from the dispersion relation using the frequency difference of wave-vector pairs with difference in accumulated phase of $2\pi$: $\Delta k=2\pi/D$.
The agreement between the experimentally obtained periodicities and both calculation approaches confirms the validity of the modeling approach. In the supplementary material the back-action on the input antenna ($S_{11}$) is presented along a short discussion on purely spin-wave coupling induced ripples in the passband.

\subsection{Impact of antenna layout}\label{sec:Antenna}

\begin{figure*}[htb]
\includegraphics{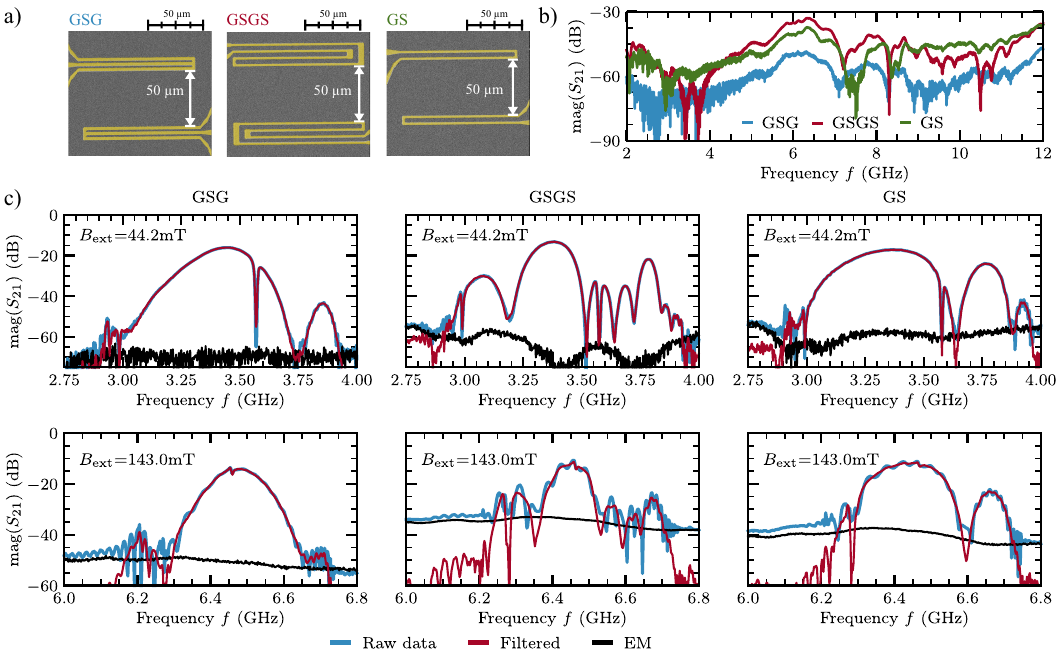}
\caption{ 
a) Colored scanning electron microscope images of all three types of antennas investigated in this work. b) Off-resonant transmission spectra of the EM acquired for all antennas and zero external magnetic field. Depending on antenna layout strong differences in EM level arise. c) Measured and filtered transmission magnitudes of for all antenna types at three different external field values along with the off-resonant EM magnitude. For all antennas the timegated signal, excluding the first $\SI{2.5}{\nano\second}$ can remove the ripples from the transmission spectra. The different shape of transmission spectra is attributed to the differences in antenna field wave-vector spectra. 
\label{fig:AntennaDependence}
}
\end{figure*} 

After confirming the origin and formation of the observed passband ripples, the influence of the antenna layout is considered. On the basis of the interference model, the modulation is strongly dependent on the magnitude of the EM coupling of both antennas and thus also on the shape of the antennas. For this reason, three different antenna layouts, shown in Fig. \ref{fig:AntennaDependence} a), are compared with regard to the magnitude of the EM. 
The measured EM at zero external magnetic field, is shown in Fig. \ref{fig:AntennaDependence} b). Across the frequency range, the GSG antenna (blue) exhibits the lowest level of crosstalk, while  either the GS (green) or GSGS (red) antenna alternately show the strongest EM contribution depending on frequency. Consequently, the formation of  passband ripples is predicted to vary depending on the antenna type and the chosen operation frequency. Please note that all antenna types exhibit an increased EM between $\SI{5.5}{\giga\hertz}$ and $\SI{7}{\giga\hertz}$, which is attributed to direct EM between the contact lines used for RF-probe connection.\\
Figure \ref{fig:AntennaDependence} c) presents the measured spin-wave transmission spectra in distinct frequency ranges for the three antennas under investigation. The sharp dips evident in the spectra of the GS and GSG antennas are not effects of interference; rather, they are attributed to gaps in the excitation spectrum arising from the hybridisation of the spin-wave surface mode with higher order thickness modes. Within the lower frequency range, the electrical crosstalk is found to be minimal across all antenna types. Consequently, no pronounced ripples are evident in the raw spectra for all antennas, which exhibit near-perfect overlap with the spectra filtered using the timegating method from the preceding section. However, for frequencies of approximately $\SI{6}{\giga\hertz}$, the EM levels are higher for all antennas. Consequently, all three antennas, including the GSG antenna, exhibit the previously identified ripples in the raw data, which can be eliminated via timegating filtering. The crosstalk strengths depicted in Fig. \ref{fig:AntennaDependence} b) reveal the strongest modulation for the meander antenna, while it is least evident for the GSG antenna. This comparison clearly underlines that the antennas must be adapted to the corresponding operation frequency range in order to achieve the best possible suppression of modulation ripples.\\

\subsection{Power dependence}

\begin{figure*}[htb]
\includegraphics{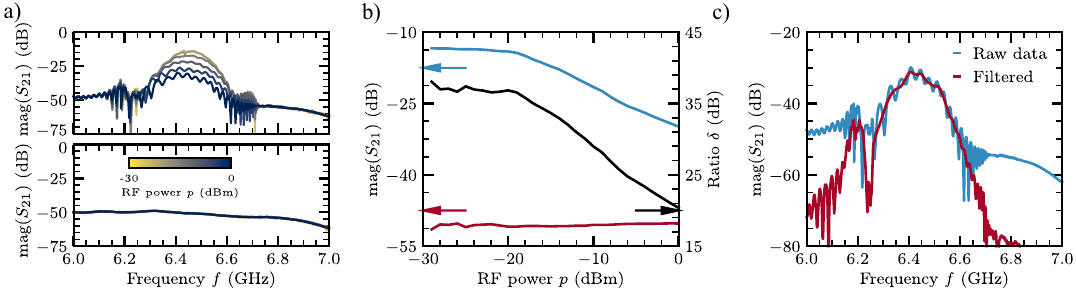}
\caption{a) Measured transmission spectra for the GSG antenna at an external field of $\SI{142}{\milli\tesla}$ (top panel) and $\SI{0}{\milli\tesla}$ (bottom panel) for varying input power. b) Extracted transmission magnitudes as function of input power for maximum spin-wave transmission and the corresponding transmission in the off-resonant case. The ratio of both magnitudes is shown in black. c) Measured transmission spectrum for an input RF power of $\SI{0}{\deci\bel}$ before and after filtering the influence of electric crosstalk in the first $\SI{2.5}{\nano\second}$ of the timetrace. \label{fig:PowerDependence}
}
\end{figure*} 

In the previous measurements, the input power was always kept in the linear transmission regime. Within this regime, the ratio of spin-wave to EM contribution is constant with respect to variation of RF input power.  As the operation in the nonlinear regime is desired for some use cases, i.e. frequency-selective limiting \cite{Davidkova}, the impact on the passband ripples is explored in the following using the GSG antenna. In the nonlinear regime, additional spin wave losses occur due to scattering processes in the spin-wave system, leading to a decrease in the relative spin-wave magnitude \cite{Ich,Davidkova}, while the EM scales linear. Accordingly, a relative weakening is expected for the spin-wave transmission in the experiment, while no change is expected for the EM, as is observed in Fig. \ref{fig:PowerDependence} a). Here it is important to note, that in the experiment the transmission magnitude is measured relative to the input. A linear power dependence therefore translates to no observed changes in transmission signal, while the spin-wave attenuation leads to a decreased transmission. Importantly, the magnitude of the passband ripples is observed to increase with input power if the spin-wave transmission enters the nonlinear regime.
This can be understood with the observation that the relative magnitude ratio between spin waves and EM gets closer to 1 and thus the modulation by the interference increases. For a better estimation, the extracted signal magnitudes are plotted at the frequency of maximum spin-wave transmission in Fig. \ref{fig:PowerDependence} b). For the magnitude of the spin waves, the influence of the electrical crosstalk was first excluded by timegating, as in section \ref{sec:Antenna}. The spin waves (blue) show an initially constant power level in the linear regime, which decreases towards high powers. For the EM (red), this curve is constant for all powers. Accordingly, the ratio of the two signal levels (black) is significantly lower for high powers than for low powers, leading to a strong increase in modulation ripples.
To check whether the observed modulation for high powers is induced by the described mechanism, Fig. \ref{fig:PowerDependence} c) shows the measured spectrum at $\SI{0}{dBm}$ before and after the timegating filter. As for the linear measurements, the modulation can be filtered out by excluding the EM. This observation  shows  the influence of the EM is always present and can effectively only be minimized by a large magnitude ratio. This behavior is consistent with previous studies \cite{Davidkova,SongSingle}, also observing similar spectral features, underlining the importance of improving the efficiency of microscale spin-wave devices while reducing interference effects.

\section{Conclusion}
In this work, we presented a simple analytic description of the interaction between spin-wave transmission and EM inside integrated micron-sized spin-wave transducers. The interference of both is found to induce passband ripples as well as distortions in phase and group delay, that depend on the relative magnitude between both contributions. The ratio is observed to be power-dependent due to the easily achievable nonlinear saturation of spin-wave transmission magnitude at high power levels. 
It is important to note here this interference effect has been studied only in continuous wave (CW) operation. For pulsed excitation, with a pulse duration shorter than the propagation time of the spin waves, there is no temporal overlap and therefore no interference between the EM and the spin-waves excited by the same pulse. However, an interference between pulses in a pulse train might occur. In CW operation, suppression of the interference is possible by reduction of the EM magnitude, i.e. by choosing antenna layouts with reduced EM. The work lays an important foundation for the characterization and reduction of unwanted passband ripple and phase distortion in integrated magnonic devices on the micron scale.

\section{Supplementary material}
The supplementary material provides further information about the YIG characterization and the antenna structures. Additionally, calculations varying the parameters forming the passband ripples along explanations of the back-action on the driving antenna and an intrinsic source of the passband ripples are included.

\section{Acknowledgements}
The authors thank Johan Vanbellingen and the imec cleanroom technical support team for their assistance with device fabrication.\\
Financial support by the EU Horizon Europe research and innovation program within the projects “MandMEMS” (Grant No. 101070536)  and “SPIDER” (Grant No. 101070417) is gratefully acknowledged. 
\bibliography{refs}
\end{document}